\input harvmac
\yearltd=\year

 
\noblackbox 

\input epsf.tex

\overfullrule=0mm

\newcount\figno
\figno=0
\def\fig#1#2#3{
\par\begingroup\parindent=0pt\leftskip=1cm\rightskip=1cm\parindent=0pt
\baselineskip=11pt
\global\advance\figno by 1
\midinsert
\epsfxsize=#3
\centerline{\epsfbox{#2}}
\vskip 12pt
{\bf Fig. \the\figno:} #1\par
\endinsert\endgroup\par
}
\def\figlabel#1{\xdef#1{\the\figno}}
\def\encadremath#1{\vbox{\hrule\hbox{\vrule\kern8pt\vbox{\kern8pt
\hbox{$\displaystyle #1$}\kern8pt}
\kern8pt\vrule}\hrule}}

\def\tr{{\rm tr} }

\def\a{\alpha}
\def\b{\beta}
\def\g{\gamma}
\def\l{\lambda}
\def\t{\theta}
\def\p{\partial}
\def\lb{\bar\lambda}

\def\cA{{\cal A}}
\def\cD{{\cal D}}
\def\cF{{\cal F}}

\def\cN{{\cal N}}
\def\cV{{\cal V}}
\def\calN{{\cal N}}
\def\calO{{\cal O}}

\lref\HoweWJ{
  P.~S.~Howe, K.~S.~Stelle and P.~C.~West,
  ``A Class Of Finite Four-Dimensional Supersymmetric Field Theories,''
  Phys.\ Lett.\  B {\bf 124}, 55 (1983).
}
\lref\HoweTM{
  P.~S.~Howe, K.~S.~Stelle and P.~K.~Townsend,
  ``The Relaxed Hypermultiplet: An Unconstrained N=2 Superfield Theory,''
  Nucl.\ Phys.\  B {\bf 214}, 519 (1983).
}

\lref\BerkovitsBT{
  N.~Berkovits,
  ``Pure spinor formalism as an N = 2 topological string,''
  JHEP {\bf 0510}, 089 (2005)
  [arXiv:hep-th/0509120].
}
\lref\BerkovitsVC{
  N.~Berkovits,
  ``New higher-derivative R**4 theorems,''
  Phys.\ Rev.\ Lett.\  {\bf 98}, 211601 (2007)
  [arXiv:hep-th/0609006].
}
\lref\AisakaYP{
  Y.~Aisaka and N.~Berkovits,
  ``Pure Spinor Vertex Operators in Siegel Gauge and Loop Amplitude
  Regularization,''
  arXiv:0903.3443 [hep-th].
}
\lref\BerkovitsVI{
  N.~Berkovits and N.~Nekrasov,
  ``Multiloop superstring amplitudes from non-minimal pure spinor formalism,''
  JHEP {\bf 0612}, 029 (2006)
  [arXiv:hep-th/0609012].
}

\lref\DixonTalk{L.~Dixon,   {\sl 	Multi-Loop Amplitudes with Maximal Supersymmetry  }  Talk   given   at  the   meeting
``International workshop on gauge and string amplitudes'', 30 March - 3 April 2009
IPPP, Durham, UK. \hfill\break
Z.~Bern, J.~J.~M.~Carrasco, L.~Dixon, R.~Roiban and H.~Johansson, {\sl
To appear.}}

\lref\PeetersQJ{
  K.~Peeters, P.~Vanhove and A.~Westerberg,
  ``Supersymmetric higher-derivative actions in ten and eleven dimensions,  the
  associated superalgebras and their formulation in superspace,''
  Class.\ Quant.\ Grav.\  {\bf 18}, 843 (2001)
  [arXiv:hep-th/0010167].
}
\lref\DrummondEX{
  J.~M.~Drummond, P.~J.~Heslop, P.~S.~Howe and S.~F.~Kerstan,
  ``Integral invariants in N = 4 SYM and the effective action for  coincident
  D-branes,''
  JHEP {\bf 0308}, 016 (2003)
  [arXiv:hep-th/0305202].
}

\lref\GreenYU{
  M.~B.~Green, J.~G.~Russo and P.~Vanhove,
  ``Ultraviolet properties of maximal supergravity,''
  Phys.\ Rev.\ Lett.\  {\bf 98}, 131602 (2007)
  [arXiv:hep-th/0611273].
}
\lref\HoweUI{
  P.~S.~Howe and K.~S.~Stelle,
  ``Supersymmetry counterterms revisited,''
  Phys.\ Lett.\  B {\bf 554}, 190 (2003)
  [arXiv:hep-th/0211279].
}

\lref\BossardSY{
  G.~Bossard, P.~S.~Howe and K.~S.~Stelle,
  ``The ultra-violet question in maximally supersymmetric field theories,''
  Gen.\ Rel.\ Grav.\  {\bf 41}, 919 (2009)
  [arXiv:0901.4661 [hep-th]].
}

\lref\MandelstamCB{
  S.~Mandelstam,
  ``Light Cone Superspace And The Ultraviolet Finiteness Of The N=4 Model,''
  Nucl.\ Phys.\  B {\bf 213}, 149 (1983).
}
\lref\BergshoeffDE{
  E.~A.~Bergshoeff and M.~de Roo,
  ``The Quartic Effective Action Of The Heterotic String And Supersymmetry,''
  Nucl.\ Phys.\  B {\bf 328}, 439 (1989).
}
\lref\deRooZP{
  M.~de Roo, H.~Suelmann and A.~Wiedemann,
  ``The Supersymmetric effective action of the heterotic string in
  ten-dimensions,''
  Nucl.\ Phys.\  B {\bf 405}, 326 (1993)
  [arXiv:hep-th/9210099].
}
\lref\TseytlinFY{
  A.~A.~Tseytlin,
  ``On SO(32) heterotic - type I superstring duality in ten dimensions,''
  Phys.\ Lett.\  B {\bf 367}, 84 (1996)
  [arXiv:hep-th/9510173].
}
\lref\TseytlinBI{
  A.~A.~Tseytlin,
  ``Heterotic - type I superstring duality and low-energy effective actions,''
  Nucl.\ Phys.\  B {\bf 467}, 383 (1996)
  [arXiv:hep-th/9512081].
}
\lref\BachasMC{
  C.~Bachas, C.~Fabre, E.~Kiritsis, N.~A.~Obers and P.~Vanhove,
  ``Heterotic/type-I duality and D-brane instantons,''
  Nucl.\ Phys.\  B {\bf 509}, 33 (1998)
  [arXiv:hep-th/9707126].
}
\lref\StiebergerWK{
  S.~Stieberger and T.~R.~Taylor,
  ``Non-Abelian Born-Infeld action and type I - heterotic duality. II:
  Nonrenormalization theorems,''
  Nucl.\ Phys.\  B {\bf 648}, 3 (2003)
  [arXiv:hep-th/0209064].
}
\lref\ZhengUM{
  Z.~J.~Zheng, J.~B.~Wu and C.~J.~Zhu,
  ``Two-loop superstrings in hyperelliptic language. I: The main results,''
  Phys.\ Lett.\  B {\bf 559}, 89 (2003)
  [arXiv:hep-th/0212191].
 Z.~J.~Zheng, J.~B.~Wu and C.~J.~Zhu,
  ``Two-loop superstrings in hyperelliptic language. II: The vanishing of the
  cosmological constant and the non-renormalization theorem,''
  Nucl.\ Phys.\  B {\bf 663}, 79 (2003)
  [arXiv:hep-th/0212198].
Z.~J.~Zheng, J.~B.~Wu and C.~J.~Zhu,
``Two-loop superstrings in hyperelliptic language. III: The four-particle
  amplitude,''
  Nucl.\ Phys.\  B {\bf 663}, 95 (2003)
  [arXiv:hep-th/0212219].
}
\lref\DHokerJC{
  E.~D'Hoker and D.~H.~Phong,
  ``Two-Loop Superstrings VI: Non-Renormalization Theorems and the 4-Point
  Function,''
  Nucl.\ Phys.\  B {\bf 715}, 3 (2005)
  [arXiv:hep-th/0501197].
}
\lref\BerkovitsNG{
  N.~Berkovits and C.~R.~Mafra,
  ``Equivalence of two-loop superstring amplitudes in the pure spinor and  RNS
  formalisms,''
  Phys.\ Rev.\ Lett.\  {\bf 96}, 011602 (2006)
  [arXiv:hep-th/0509234].
}
\lref\BerkovitsDF{
  N.~Berkovits,
  ``Super-Poincare covariant two-loop superstring amplitudes,''
  JHEP {\bf 0601}, 005 (2006)
  [arXiv:hep-th/0503197].
}

\lref\BernUG{
  Z.~Bern, L.~J.~Dixon, D.~C.~Dunbar, M.~Perelstein and J.~S.~Rozowsky,
  ``On the relationship between Yang-Mills theory and gravity and its
  implication for ultraviolet divergences,''
  Nucl.\ Phys.\  B {\bf 530}, 401 (1998)
  [arXiv:hep-th/9802162].
}

\lref\BernCT{
  Z.~Bern, J.~J.~M.~Carrasco, H.~Johansson and D.~A.~Kosower,
  ``Maximally supersymmetric planar Yang-Mills amplitudes at five loops,''
  Phys.\ Rev.\  D {\bf 76}, 125020 (2007)
  [arXiv:0705.1864 [hep-th]].
}
\lref\BrinkWV{
  L.~Brink, O.~Lindgren and B.~E.~W.~Nilsson,
  ``The Ultraviolet Finiteness Of The N=4 Yang-Mills Theory,''
  Phys.\ Lett.\  B {\bf 123}, 323 (1983).
}

\lref\GreenMN{
  M.~B.~Green, J.~H.~Schwarz and E.~Witten,
  ``Superstring Theory. Vol. 2: Loop Amplitudes, Anomalies And Phenomenology,''
{\it  Cambridge, Uk: Univ. Pr. ( 1987) 596 P. ( Cambridge Monographs On Mathematical Physics)}
}
\lref\GrassiFE{
  P.~A.~Grassi and P.~Vanhove,
  ``Higher-loop amplitudes in the non-minimal pure spinor formalism,''
  JHEP {\bf 0905}, 089 (2009)
  [arXiv:0903.3903 [hep-th]].
}

\lref\BernNH{
  Z.~Bern, J.~S.~Rozowsky and B.~Yan,
  ``Two-loop four-gluon amplitudes in N = 4 super-Yang-Mills,''
  Phys.\ Lett.\  B {\bf 401}, 273 (1997)
  [arXiv:hep-ph/9702424].
}

\Title{\vbox{\hbox{DAMTP-2009-41, IPHT-T-09-086, IHES/P/09/35, ICCUB-09-221}}}
{\vbox{ 
\centerline{Non-renormalization conditions for four-gluon}
\medskip
\centerline{  scattering in supersymmetric string and field theory}
}} 
\vskip-1cm\centerline{\bf Nathan Berkovits${}^a$, Michael B. Green${}^b$,
Jorge G. Russo${}^{c,d}$, Pierre Vanhove${}^{e,f}$} 
\smallskip
\centerline{${}^a$ \sl Instituto de F\'\i sica Te\'orica, 
UNESP-Universidade Estadual
Paulista,} 
\centerline{\sl Caixa Postal 70532-2, S\~ao Paulo, SP 01156-970, Brasil} 
\centerline{${}^b$ \sl DAMTP,
Wilberforce Road, Cambridge CB3 0WA, UK}

\centerline{${}^c$ \sl Instituci\'o Catalana de Recerca i Estudis
Avan\c cats (ICREA)}
\centerline{${}^d$ \sl Departament ECM and Institut de Ciencies del Cosmos, Facultat de Fisica,}
\centerline{\sl  University de Barcelona, 
 Av. Diagonal, 647, Barcelona 08028 SPAIN}
\centerline{${}^e$ \sl IHES,  Le  Bois-Marie, 35  route de  Chartres,
F-91440 Bures-sur-Yvette, France}
\centerline{${}^f$ \sl Institut de Physique Th\'eorique, CEA/Saclay,   F-91191 Gif-sur-Yvette,
France}
\medskip
\centerline{\bf Abstract} {\baselineskip =8pt} 

The constraints imposed
by maximal supersymmetry on multi-loop contributions to the scattering
of four open superstrings in the  $U(N)$ theory are examined by use of
the pure  spinor formalism.  The double-trace term 
$k^2\,t_8(\tr F^2)^2$ (where $k  $ represents an external momentum and
$F$ the Yang--Mills field strength)  only receives contributions from $L\leq
2$ (where $L$ is the loop number)
while the
single-trace term  $k^2\,t_8(\tr F^4)$ receives
contributions from  all $L$.  These statements are verified up to
$L=5$, but arguments based on supersymmetry suggest they extend to all $L$.  
This explains
why   the   single-trace  contributions   to   low  energy   maximally
supersymmetric  Yang--Mills field  theory  are more  divergent in  the
ultraviolet than the double-trace contributions. 
 We also comment further on the constraints on closed string amplitudes
and their implications for ultraviolet divergences in $\cN=8$ supergravity.

\Date{\the\day/\the\month/\the\year}

\newsec{Introduction}

Supersymmetry  imposes   crucial  constraints  on   the  structure  of
scattering  amplitudes  in   supersymmetric  gauge  and  gravitational
theories,  which  generally  leads  to  a  moderation  of  ultraviolet
divergences. These  constraints are particularly  strong for maximally
supersymmetric  theories,   which  are  difficult   to  analyse  using
conventional superspace  techniques due to the absence  of an off-shell
superspace formalism.   However, it is possible to  analyse such field
theory supersymmetry  constraints by considering the  low energy limit
of  the  corresponding  open   or  closed  superstring  theories.   In
particular, the pure spinor formalism \refs{\BerkovitsBT,\BerkovitsVI}
is a framework for constructing multi-loop string theory amplitudes in
a  manner  that  preserves  all the  space-time  supersymmetries.   An
example of constraints obtained in this manner comes from the analysis
of multi-loop contributions to  the four-graviton amplitude in type II
superstring theory  \refs{\BerkovitsVC}.  These constraints  imply that
interactions of the form  $\partial^{2k}\, R^4$ (where $R^4$ denotes a
particular  contraction of  four Riemann  curvatures) do  not  get any
perturbative  contributions beyond  $k$ loops  in  the ten-dimensional
theory, at least for $k\le  6$.  These conditions follow from the fact
that  interactions  with $k  <  6$ are  $F$-terms  that  are given  by
integrals  over a  fraction of  the full  32-component  superspace.  A
striking consequence of this  that follows on purely dimensional grounds
is that ultraviolet  divergences should be absent up to  at least nine loops
in four-dimensional ($D=4$)  $\cN=8$ supergravity \refs{\GreenYU}.  By
contrast,  analyses of counterterms  that exploit  less than  the full
${\cal     N}=8$      supersymmetry     give     weaker     conditions
\refs{\HoweUI,\BossardSY}.

The main purpose of this paper  is to extend  these considerations to
open string theory  and, hence, to its low  energy limit --- maximally
supersymmetric  Yang--Mills  (SYM)   theory.

 \subsec{General properties of the four-gluon amplitude}
 
 For simplicity we  will consider the case of  open strings scattering
 on  $N$   coincident  $Dp$-branes,  for  which   the  world-sheet  is
 orientable and which corresponds to  a $U(N)$ gauge theory in the low
 energy field theory  limit.  It has long been  known that ultraviolet
 divergences are absent  in maximally supersymmetric Yang--Mills field
 theory in dimensions  $D\leq 4$ to all orders  in perturbation theory
 --- one  does not  even need  to exploit  the full  power  of maximal
     supersymmetry to argue that the theory is UV
finite~\refs{\MandelstamCB,\BrinkWV,\HoweWJ,\HoweTM}.   Indeed,   there  are   finite  $\cN=2$  and   $\cN=1$  super
Yang--Mills  theories.  It is  sufficient to  know that  the dimension
four operator $t_8 F^4$ factors out  in the sum of Feynman diagrams at
every   order  in   perturbation  theory  (where  $t_8$  is  a standard  eight-index  tensor reviewed in appendix 9.A. 
of~\refs{\GreenMN}  that contracts  the
space-time indices in $F^4$ while the contraction of the gauge indices
will be  discussed later).  It follows by  simple dimensional analysis that the perturbative contributions are ultraviolet finite at each order  (all $L$) in dimensions $D\le
4$.  However,  the situation is  better than that because  the CP-even
$t_8 F^4$ is related by supersymmetry to the CP-odd anomaly cancelling
term $B\wedge F^4$ in ten dimensions, which is expected to be one-loop
exact~\refs{\BergshoeffDE,\deRooZP,\TseytlinFY,\TseytlinBI,\BachasMC,\StiebergerWK,\ZhengUM,\DHokerJC,\BerkovitsNG,\BerkovitsDF}.
This  means that  the  $L>1$ contributions  to  the string  scattering
amplitude must have a low energy limit that behaves as $s^{\gamma_L}\,
t_8\, F^4$  with $\gamma_L \ge 1$,  so that the  prefactor contains at
least two  extra powers of momentum\foot{The  factor of $s^{\gamma_L}$
in  this expression,  and all  those that  follow, is  intended to
indicate the power of  Mandelstam invariants -- the precise expression
involves a function of $s,t,u$ with a detailed structure that will not
concern us here.  }.  We may  interpret this contribution as a term of
the form  $\partial^{2\gamma_L}\, t_8\,  F^4$ in the  effective action
and in the following we will  often pass between the amplitude and the
effective action without comment.  In fact, there are indications that
$\gamma_L =1$ for $L>1$ from
 direct  perturbative  evaluations  of  the  four  gluon
amplitude      in       maximally      supersymmetric      Yang--Mills
theory~\refs{\BernUG,\BernCT,\DixonTalk} or by $\cN=3$ superspace
arguments in four dimensions~\refs{\HoweUI}.  Assuming $\gamma_L=1$, it
is easy  to see using  dimensional analysis that an  $L$-loop amplitude
with a prefactor of $s\, t_8 F^4$ is ultra-violet finite in dimensions
$D< 4+6/L$.

Although extended supersymmetry  determines the dynamical prefactor to
be of  the form  $s^{\gamma_L}\, t_8 F^4$~\refs{\PeetersQJ},  there is
also  a dependence  on  the gauge  group  and on  the string  coupling
constant $g_s$,  which is related to the  Yang--Mills coupling $g_{\rm
YM}$ by $g_s  = g_{\rm YM}^2/4\pi$.  For example,  for the gauge group
$U(N)$ (which is the simplest example) there are two independent group
theory  structures in  the  field theory  four-gluon  amplitude ---  a
single trace term $t_8 \tr(F^4)$  and a double-trace $t_8 (\tr F^2)^2$
term, where we are  now taking $F$ to be an $N\times  N$ matrix in the
defining representation of  $U(N)$ and $\tr$ denotes a  trace on these
indices.  Contributions to the  amplitude of the general form $s^{\gamma_L}\,
t_8  \tr(F^4)$ and  $s^{\beta_L}\, t_8(\tr  F^2)^2$ will  be  constrained by
supersymmetry in  different manners at  a given order  in perturbation
theory (i.e.,  for a given power  of $g_s$).  Our aim  is to determine
the values of  $\gamma$ and $\beta$ by considering the low  energy limit of the
four-gluon amplitude in open superstring theory.

Certain properties  of the  open superstring four-gluon  amplitude are
well known.  For  example, at tree-level ($L=0$) the  world-sheet is a
disk with all vertex operators describing the external states coupled to the boundary.  In
 this case, in the low energy limit only a single-trace
$g_s^{-1}\,   t_8\,  \tr(F^4)$  term   contributes.  
 For   $L=1$  the
world-sheet is an annulus, which  has two boundaries.  When all vertex
operators  are attached to  a single  boundary there  is a  low energy
contribution that behaves as  $N\, t_8\,\tr(F^4)$ (where the factor of
$N$ arises  from a  trace over  the free boundary).   There is  also an
$L=1$ contribution when  there is a pair of  vertex operators attached
to each  boundary, which reduces in  the low energy limit  to terms of
the  form  $t_8(\tr F^2)^2$.   These are
$F$-terms   that  are  protected  from   higher-loop   ($L>1$)  quantum
corrections. By an $F$-term, we mean a term which cannot be expressed
as an integral over 16 $\t$'s of a gauge-invariant integrand. On the
other hand, a $D$ term is 
a term which can be expressed
as an integral over 16 $\t$'s of a gauge-invariant integrand. 
We  will confirm, using string  loop calculations, that
$t_8\tr(F^4)$   and    $t_8(\tr   F^2)^2$  are $F$-terms that satisfy  the expected $L>1$
non-renormalization properties, at least up to $L=5$.

The
 next term that arises in the low energy expansion is the dimension
 ten operator
$s\, t_8\tr(F^4)$, which  can be expressed as a  term in the effective
action  given by  a  superspace integral,  of  the form  $I_{1/4}=\int
d^{10}x \int d^{16}\theta \, \theta^8\, \tr(W^4)$, where $W_\a$ is the
gaugino  superfield  ($\a$  is   a  space-time  spinor  index).   This
expression is schematic since we  have not specified how to factor out
the eight  powers of  $\theta$ in a  covariant manner.   However, this
term is a ``fake'' $F$-term because  it can be rewritten
as a $D$-term, at least  after compactification to dimension $D < 10$.
For example, in four dimensions it has the form $I_{1/4}=\int d^4x\int
d^{16}\theta \, K$, where $K=\tr(\varphi^i\varphi_{i})$ is the Konishi
operator and  $\varphi_i$ is the scalar  superfield in the  {\bf 6} of
the R-symmetry group $SO(6)\sim SU(4)$ \refs{\DrummondEX,\BossardSY}.
The  fact  that  this  integral  contains  $\partial^2\,t_8  \tr(F^4)$
follows  from  the  nonlinear  completion of  the  linearised  $\cN=4$
superfield.    Our string calculations will confirm  that $\partial^2\, t_8\tr(F^4)$ is not protected against quantum
corrections and receives contributions for all $L>1$,
as  expected for
$D$-terms.  We will also show from string theory that the corresponding 
double-trace   contribution  $s\,   t_8   (\tr  F^2)^2 
\sim \partial^2\, t_8  (\tr F^2)^2$ 
only receives  perturbative  corrections  up to two  loops ($L\leq 2$),  
as expected for a protected  $F$-term 
that cannot be written as a $D$-term.

Furthermore, the dimension-twelve double-trace
operator $s^2\, t_8(\tr F^2)^2$ can be
expressed as a term in  the effective action, $I_{1/8}=\int d^4x\int
d^{16}\theta\, \theta^4\, (\tr W^2)^2$, which 
can also be rewritten as a $D$-term using 
 $ \int d^4x  \int d^{16}\theta K^2$ and $\int  d^4x \int d^{16}\theta
 \,T^2$ where $T_{AB}$ is the symmetric traceless supercurrent~\refs{\DrummondEX}.
 An analogous description of this $D$-term 
in terms of a scalar superfield should exist in all  dimensions with $D <10$.
We will again use the string theory multi-loop amplitude prescription
to argue that this
double-trace contribution $s^2\, t_8 (\tr F^2)^2 \sim \partial^4\, t_8
(\tr F^2)^2$ receives perturbative
corrections at all loop orders, as expected for a $D$-term.

\subsec{ Organization  of the paper}

In section~2  we will
review the construction of  scattering amplitudes on orientable planar world-sheets with
$L$ open-string  loops, in the  non-minimal version of the  pure spinor
formalism.  This describes open strings moving in $D$ 
dimensions and scattering on $N$ coincident $Dp$-branes (where $D=p+1$).  The  amplitude for the  scattering of four gluons  will be
considered  in section~3,  where we  will  highlight the  distinction
between contributions from single trace and double trace terms.
We will also make a short comment that supports the arguments concerning
$F$-terms  in graviton scattering in type II closed string theories in
\refs{\BerkovitsVC}.  In section~4 we will discuss the contributions of world-sheet handles, which are associated with the coupling of the closed-string (i.e., gravitational) sector and generate contributions suppressed by $O(1/N^2)$ relative to the zero-handle terms.
In section~5 we  will show how the  structure of
these string  theory expressions  explains the pattern  of ultraviolet
divergences in  maximally supersymmetric $U(N)$  Yang--Mills theory at
$L$  loops in  $D$  dimensions.  
Whereas the  onset of ultraviolet  divergences of the  single-trace term $s\,t_8 \tr(F^4)$ to the  $L$-loop amplitude occurs in dimensions $D= 4 + 6/L$, our results imply that the dimensional
dependence of the double-trace  terms is different.  Given the overall
prefactors described above, together  with dimensional analysis we will see
that  ultraviolet divergences for the double-trace terms  arise when:\break
\noindent $\bullet$\  $D= 8$ for the $L=1$ term $ t_8 (\tr F^2)^2$;

\noindent $\bullet$\ $D= 7$ for
the $L=2$ term $s\,t_8 (\tr F^2)^2$;

\noindent $\bullet$\ $D= 4+8/L$ for
the term $s^2\,t_8 (\tr F^2)^2$ at $L\geq3$.

\noindent This  explains the  apparent
puzzles  that have  arisen  in the  explicit multi-loop  calculations,
where the double trace $t_8\,(\tr F^2)^2$ contribution to the $L$-loop
counter-term is absent in dimensions $D=4+6/L$ for $L=3$ and $L=4$ (as
reviewed in~\refs{\DixonTalk}).    
 Finally, we will summarize our results in section~6 and  make some preliminary comments concerning
higher-point gluon amplitudes.

\newsec{Open-string scattering amplitudes in the pure spinor formalism}

The functional integral that defines the scattering amplitude with $M$
external  massless  ground states  (``gluons'')  includes  a sum  over
boundaries, handles  and cross-caps (for  theories with non-orientable
world-sheets).   Recall  that   a  world-sheet  with  $M$  open-string
vertices and  with $B$ boundaries,  $H$ handles and $C$  cross-caps is
weighted  with a  factor $g_s^{-\chi}$, where  $\chi$, the
Euler  number, is given  by $\chi=2-  2H-B-C$ and  where $g_s$  is the
string coupling. When  describing the scattering of open  strings on a
collection of $Dp$-branes we need to consider orientable world-sheets,
so   that  $C=0$.   In   addition  we   will  initially   neglect  the
contributions of world-sheets  with handles so we will  set $H=0$.  
We are  interested in taking  the limit in  which gravity
decouples  from  Yang--Mills and  handles  describe the  gravitational
contributions.   However,  handles  also  contribute  finite  residual
pieces to pure super-Yang-Mills amplitudes.  These  contributions are
suppressed by  at least two powers  of $N$ in the large  $N$ limit (since
one handle  takes the place  of two free  boundaries).  So we  will be
interested  initially  in the  $L=B-1$  -  loop  oriented open  string
corrections with $\chi=1-L$. In the low-energy limit, these open superstring
amplitudes describe $U(N)$ super-Yang-Mills amplitudes to leading order in
$1/N^2$.
  
A general property of  these open-string amplitudes is the possibility
of  divergences  associated  with   closed  strings  coupling  to  the
$Dp$-brane via  the world-sheet  boundaries.  The simplest  example is
the single-trace contribution at one loop, $L=1$ ($B=2$), which may be
viewed  as a cylinder  carrying zero  $(p+1)$-dimensional longitudinal
momentum and with both boundaries  fixed at the same transverse point.
The potential divergence arises from the massless state propagating in
the cylinder, which gives a contribution proportional to
\eqn\emassless{
\lim_{x\to 0}\int^{x^{-1}} d^{9-p} q^\perp (q^\perp)^{-2}\,,
}
where $q^\perp$ is  the transverse momentum in the  cylinder, which is
integrated  in order  to fix  the  boundaries at  the same  transverse
positions.  This  expression diverges for $p  \ge 7$,  which indicates
that  the gravitational back  reactions of  $Dp$-branes with  $p\ge 7$
cannot  be  neglected.   In   the  following,  we  will  restrict  our
considerations to  the situation in which  these gravitational effects
can  be ignored,  so we  will be  considering Yang--Mills  in $D  < 8$
dimensions.

\subsec{The multi-loop functional integral for world sheets with $B$ boundaries}

The $M$-gluon amplitude is expressed as a sum of terms in which the $M$ vertex operators are partitioned among the $B$ boundaries in all possible ways and there is a sum over the order of the operators attached to each boundary.  This gives
\eqn\eAmpone{
A_L=   \sum_{\rm
orderings} 
g_s^{B-2}\,  \tr(T_{a^{(1)}_1}  \cdots
T_{a^{(1)}_{n_1}})  \cdots \tr(T_{a^{(B)}_1} \cdots T_{a^{(B)}_{n_B}})
\, \cA_L^{(a^{(1)}_1 \cdots a^{(1)}_{n_1})\cdots ( a^{(B)}_1 \cdots a^{(B)}_{n_B})}
}
where $\{n_r\}$ is the number of vertex operators attached to each boundary labelled $r =1, \dots, B$ and 
$\sum_{r=1}^B  n_r = M$. The quantity   $\cA_L^{(a^{(1)}_1 \cdots a^{(1)}_{n_1})\cdots ( a^{(B)}_1 \cdots a^{(B)}_{n_B})}$ is the
 colour-ordered partial amplitude.  The sum is 
over all partitions of the $M$ vertex operators on the $B$ boundaries, including all possible orderings of the operators on each boundary.   The gauge group generators, $T_{a_i}$ are  $N\times N$ matrices with indices in the defining representation of the gauge group, $U(N)$
 (which is the Chan--Paton prescription).
Such a surface has $3B =  3L+3$ real moduli for $L>1$ (and one modulus
for  $L=1$).  Note  that  there is  a  factor of  $\tr(1)=N$ for  each
boundary that has no vertex operators attached, leading to an overall 
 factor of $N^{B_f}$ (where $B_f$  is the number of free boundaries in
 a given term in the sum). 

 In the  non minimal pure  spinor formalism the prescription  for each
 colour-ordered open-string amplitude is given (for $B=L+1>2$) by 
\eqn\eAmp{\eqalign{
\cA_L^{(a^{(1)}_1 \cdots a^{(1)}_{n_1})\cdots ( a^{(B)}_1 \cdots a^{(B)}_{n_B})}
= &
\int d^{3B-6}\tau\, \Big\langle
\prod_{i=1}^{3B-6}(\mu_i|b)\,  \cN
\prod_{i=1}^M \int dt_i~ :\cV^{a_i}_i(t_i)e^{ik^i\cdot x}:\Big\rangle
}}
where $(\mu|b):=\int d^2y\, \mu_{\bar z}{}^z\, b_{zz}$ (and $\mu_{\bar
z}{}^z(\tau_{a})= g_{z\bar z}\partial g^{zz}/\partial \tau_{a}$ is the
Beltrami differential) and $\tau_{a}$ are the Teichm\"uller parameters
of  the bordered  Riemann  surface.   For a  given  colour factor  the
positions of the vertex operators ${\cal V}_i^{a_i}$ are integrated in
a given order along the boundary.

The  angular  bracket $\langle  \cdots  \rangle$  represents the  path
integral over  the matter  fields $[x^m,\theta^\a,p_\a]$ and  the pure
spinor  ghosts which consist  of left-movers,  $[\l^a, w_a,\lb_\a,\bar
w^a,      r_\a,s^\a]$,     and      right-movers,     $[\tilde{\l}^\a,
\tilde{w}_\a,\tilde{\lb}_\a,\tilde{\bar             w}^\a,\tilde{r}_\a,
\tilde{s}^\a]$,  weighted  by the  pure  spinor action  \eqn\ePathint{
\langle    \cdots   \rangle=\int    \cD^{10}    x   \cD^{16}\theta\int
[\cD\l][\cD\lb][\cD  r]\prod_{I=1}^L\int [\cD  w^I][\cD  \bar w^I][\cD
s^I] \cdots\, e^{-S_{ps}} \,,}
where the action  is (setting $2\pi\alpha'=1$)
\eqn\ePsS{\eqalign{
S_{ps}=\int d^2z\,  \bigg({1\over2} \p x^m \bar\p x_m  & + p_\a\bar\p \theta^\a +
w_\a \bar\p \l^\a+ \bar w^\a \bar\p \lb_\a+s^\a \bar\p r_\a \cr
&  + \tilde p_\a  \tilde{\p} \tilde {\theta^\a} +
 \tilde {w}_\a \tilde{\p} \tilde{\l}^\a+ \tilde{ \bar w}^\a \tilde{\p}
 \tilde{\lb}_\a+ \tilde{ s}^\a \tilde{\p} \tilde r_\a\bigg)
\,.}}
This integral generically needs to be regularised by introducing the quantity $\cN$ that will be reviewed below.

The  pure  spinor  ghosts  $\l^\a$,  $\lb_\a$ and  $r_\a$  satisfy  the
constraints
\eqn\ePSconst{
\l \g^m\l=0\,,  \qquad \lb \g^m \lb=0\,, \qquad \lb \g^m r=0\, ,
}
which implies  that they each  have eleven independent  components.  A
conformal weight zero field has  a single zero mode so $\theta^\a$ has
16 fermionic zero modes and  $r_\a$ has 11 fermionic zero modes, which
all have to  be saturated in the functional  integral.  Likewise, each
conformal weight one  field has $L$ real zero  modes.  This means that
$p_\alpha$ has  $16 L$ real zero  modes while $s^\a$  has $11L$, which
also need to be saturated.  The bosonic pure spinor ghosts $\l^\a$ and
$\lb_\a$  each  have  11  independent  components  which  need  to  be
integrated.    Singularities   in   these   integrals   need   to   be
regulated~\refs{\BerkovitsBT,\BerkovitsVI,\AisakaYP,\GrassiFE}.
Similarly,  the  conjugate  bosonic  variables, $w_\alpha$  and  $\bar
w^\alpha$, each have $11L$ components.

The integration measures are given by
\eqn\edlBerk{\eqalign{
\l^\a \l^\b\l^\g\, [\cD \l]&= (\epsilon {\cal T}^{-1})^{\a\b\g}_{k_1\cdots
k_{11}} \cD \l^{k_1}\cdots
\cD \l^{k_{11}}\cr
[\cD\lb][\cD r]&= \cD \lb_{\a_1}\wedge\cdots\wedge
\cD \lb_{\a_{11}}\,\times\, \partial_{r_{\a_1}}\wedge\cdots
\wedge\partial_{r_{\a_{11}}}\ ,
}}
where the tensor  ${\cal T}$, which  is totally antisymmetric on the $k_i$
indices and fully symmetric and 
$\gamma$-traceless on the $\a\b\g$ indices, has the form~\refs{\BerkovitsBT}
\eqn\eeT{(\epsilon{\cal                  T})^{k_1\cdots
k_{11}}_{\a\b\g}=\epsilon_{16}^{k_1\cdots k_{11}r_1\cdots r_5}\, (\g^m)_{((\a
|r_1|}\, (\g^n)_{\b  |r_2|}\, (\g^p)_{\g ))r_3}\, (\g_{mnp})_{r_4r_5}\
.
}
The integration measure of the conjugate ghosts is given by 
\eqn\edwwww{\eqalign{
\l^{\a_1}\cdots\l^{\a_8}\, [\cD w^I]&=M^{\a_1\cdots \a_8}_{m_1n_1\cdots
m_{10}n_{10}} \cD N^{m_1n_1\, I}\cdots \cD N^{m_{10}n_{10}\,I} \cD J^I\cr
[\cD \bar w^I][\cD s^I]&=\prod_{i=1}^{10} \cD\bar N_{m_in_i}^I\wedge
\cD\bar J^I \wedge \prod_{i=1}^{10} \partial_{S^I_{m_in_i}}\wedge \partial_{S^I}
}}
where 
\eqn\eM{\eqalign{
M_{m_1n_1\cdots   m_{10}n_{10}}^{\alpha_1\cdots   \alpha_8}= &{ 
(\g_{m_1n_1m_2m_3m_4})^{((\a_1\a_2} 
(\g_{m_5n_5n_2m_6m_7})^{\a_3\a_4} }\cr
& { (\g_{m_8n_8n_3n_6m_9})^{\a_5\a_6} }
{(\g_{m_{10}n_{10}n_4n_7n_9})^{\a_7\a_8))}}} }
and   $((\cdots))$   means   that   one  considers   the   symmetrised
$\gamma$-traceless part.  
The quantities 
\eqn\eNN{\eqalign{
N^{mn}&= \l \g^{mn}w\, , \qquad J=\l^\a w_\a\cr
\bar  N^{mn}&=\lb\g^{mn}\bar w  - r\g^{mn}  s\,, \qquad  \bar J=\lb\cdot
\bar w- r\cdot s\cr
S^{mn}&= \lb\g^{mn}s\,, \qquad S= \lb\cdot s\,,
}}
are conserved world-sheet currents.

The $b$-ghost is a composite quantity, which is defined to satisfy the
BRST condition $[Q,b]=T$, where $Q$ is  the BRST charge and $T$ is the
energy-momentum      tensor.        This      takes      the      form
\refs{\BerkovitsBT,\BerkovitsVI},
\eqn\nnA{
b= s \p \bar\l +{\bar \l_\a \, {\bf b}^\a \over  \l \cdot \bar \l}\,,
}
and 
\eqn\eBB{
{\bf b}^\a \equiv G^\a+ {r_\b\over \l \cdot \bar \l} H^{\a\b}+ {r_\b r_\gamma\over (\l \cdot \bar \l)^2}
K^{\a\b\gamma}+{r_\b r_\gamma r_\delta\over (\l \cdot \bar \l)^3} L^{\a\b\gamma\delta}\,,}
where
\eqn\eGHKL{\eqalign{
G^\a&\equiv {1\over2}\Pi^m(\g_m d)^\a - {1 \over 4}N_{mn}(\g^{mn}\partial \theta)^\a- {1 \over 4}J\partial\theta^\a - {1 \over 4} \partial^2 \theta^\a\cr
H^{\a\b}&\equiv {1\over 192} (\g^{mnp})^{\a\b} \,\left((d\g_{mnp}d) +4! N_{mn}\Pi_p\right) \cr
K^{\a\b\g}&\equiv {1\over16} (\g_{mnp})^{[\a\b} (\g^md)^{\g]} N^{np}\cr 
L^{\a\b\g\delta}&= {1\over128} (\g_{mnp})^{[\a\b}(\g^{pqr})^{\g\delta]} N^{nm} N_{qr}  \ .
}}
The pieces of the $b$-ghost satisfy the relations
\eqn\eBQ{\eqalign{
\{Q, G^\a\} &=\l^\a \, T\,, \qquad
\{Q, H^{\a\b}\} =\l^\a \, G^\b\,,\cr
\{Q, K^{\a\b\g}\} &=\l^\a \, H^{\b\g}\,, \qquad
\{Q, L^{\a\b\g\delta}\} =\l^\a \, H^{\b\g\delta}\,.
}}
The regulator $\cN$ in \eAmp\  given by 
\eqn\Reg{\eqalign{
\cN &=\exp{[Q,\Psi]}\cr
&    =    \exp\Big[   -    \l\cdot\lb   -    r\cdot\theta\Big]\cr
&\times\exp\Big[ -
 \sum_{I=1}^{L} \Big({1\over2} N^I_{mn} \bar N^{I\,mn}+J_I\bar J^I\Big)\Big]
 \cr
&\times\exp\Big[-\sum_{I=1}^L\, {1\over4} S^I_{mn} (d^I\g^{mn}\l) 
+ S^I (\l d^I) 
\Big]\,, \cr
}}
 In this expression we have defined the $I$'th zero mode of any  current in \eNN,   $\calO^I$,  as the integral of $\calO$ around the $I$'th $a$-cycle of the doubled open-string world-sheet, $\calO^I= \oint_{a_I} dz\, \calO$.
\subsec{The open-string vertex operator}

An open string vertex operator in \eAmp\  attached to a point $t$ on a boundary,  $\cV^{a}(t)\, e^{ik\cdot x}$, is  the $k$'th Fourier mode of the position-space superfield given by 
\eqn\eU{
\cV^a(x,\theta) =   A^{a}_\a(x,\theta)\, \partial \theta^\a+  A^{a}_m(x,\theta)\, \Pi^m + 
W^{a\,\a} (x,\theta)\, d_\a
+{1\over 2} N^{mn} \cF_{mn}^{a} (x,\theta)\,,
}
where  $\Pi^m=  \p  x^m  +  i/2 (\theta\g^m  \p  \theta)$ ,  $d_\a=
p_\a-i/2\,  (\theta\g_m)_\a \,(\p  x^m  +(\theta\g^m\p\theta)/4)$, and
$A^{a}_\a$, $A^{a}_m$, $W^{a\,\a}$  and $\cF^{a}_{mn}$ are the $\cN=1$
$D=10$ super-Yang-Mills superfields,
\eqn\eSuperfield{
A^{a}_\a(x,\theta)={1\over2}\,    (\g^m\theta)_\a   \,    a^{a}_m (x)  -{1\over3}
(\chi^a(x)\, \g_m\theta)\,(\g^m\theta)_\a -{1\over32} \, F^a_{mn}(x)\, (\g_p\theta)_\a(\theta\g^{mnp}\theta)+\cdots
,}
$a^a_m(x)$ and $\chi^{\a a}(x)$ are the gluon and gluino fields and $F^a_{mn} = \partial_{[m} a^a_{n]}$,
and 
\eqn\eEOM{\eqalign{
(\g^m)_{\a\b} A_m& = D_\a A_\b+D_\b A_\a\cr
(\g_m)_{\a\b} W^\b&= D_\a A_m-\p_m A_\a\cr
D_\a W^\b&={1\over 4}(\g^{mn})_\a{}^\b \cF_{mn}\, ,\cr
}}
where $D_\a=\p/\p\theta_\a+1/2 (\theta \g^m)_\a \p_m$ is the supersymmetric derivative.
The superfield $W^\a(x,\theta)$ takes the form
\eqn\eW{
W^{a\,\a}  =\chi^{a\,\a}  -  {1\over4}  (\g^{mn}\theta)^\a  \,  F^a_{mn}+{1\over4}
(\g^{mn}\theta)(\p_m \chi^a\g^m\theta)+ {1\over 2\cdot 4!}\,
(\g^{pq}\theta)^\a (\theta\g_q\g^{mn}\theta) \p_p F^a_{mn}+O(\theta^4)\,.
}
The first two terms in \eU\ can be expressed  by means of a normal  
coordinate expansion around $Z_0=(x_0^m,\theta_0^\a)$, giving
\eqn\edA{
 dZ^M A_M = A_M(Z_0)  dZ^M +\cF_{MN} (Z-Z_0)^M dZ^N+ \cdots 
}
The  first term  in this  expression can  be ignored  since it  can be
written as  the surface  term $d(A_M(Z_0) Z^M)$  which decouples  using the
standard canceled  propagator argument. And because  $Z-Z_0$ does not
contain $\theta$  zero modes, $\cF_{MN} (Z-Z_0)^M dZ^N+  \cdots $ only
contributes to terms in the effective action which are higher order in
derivatives than the  terms coming from $d_\a W^\a$.   This is easy to
see since  $d_\a W^\a$ can contribute  a $d_\a$ zero  mode whereas the
term $W^\a  [ (\t-\t_0) dX  - (X-X_0) d\t]_\a$ in  $\cF_{MN} (Z-Z_0)^M
dZ^N$    cannot   contribute   either    $d_\a$   or    $\t^\a$   zero
modes.  Furthermore,  the term  $N^{mn}  \cF_{mn}$  in  \eU\ does  not
contain  $d_\a$ zero modes  so it  also only  affects terms  which are
higher  order  in derivatives  than  terms  coming  from $W^\a  d_\a$.
Therefore, when analyzing the terms  in the effective action of lowest
dimension  at  a   given  genus,  one  only  needs   to  consider  the
contribution from $W^\a d_\a$ in \eU.

\newsec{The scattering of four open strings}

We will now specialise to  the scattering of four massless open-string
ground  state gluons with  momenta $k_r$  ($r=1,2,3,4$) satisfying
$k_r^2=0$.  In this case there  are two distinct contributions at each
order  in perturbation  theory, which  differ in  the way  their gauge
indices are contracted.
  One of  these  is the
single trace  term in which all  vertex operators are  attached to one
boundary,  resulting in  an  overall factor  of $N^L\,  s^{\gamma_L}\,
t_8\tr(F^4)$ in the low energy  limit of the amplitude without handles
$H=0$.  For  $L=1$ we  know $\gamma_0= \gamma_1=0$  and we  will argue
shortly that  $\gamma_L =1$ for $L>1$.  As we
will see  later in this section,  $s\,t_8 \tr(F^4)$
gets contributions from  all values of  $L>1$, as
expected for a $D$-term.  The
second kind of contribution is the double trace term, arising when the
four vertex operators are partitioned in pairs between two boundaries,
resulting in a factor $N^{L-1}\, s^{\beta_L}\,t_8 (\tr F^2)^2$.  There
is  no  such contribution  at  tree level  ($L=0$)  and  we know  that
$\beta_1=0$.  We  will see shortly  that $\beta_L= \lceil  L/2 \rceil$
for  $1<L\leq 4$ (recalling that  $\lceil x  \rceil  $ denotes  the smallest  integer
greater or equal  than $x$) again up  to at least $L=4$.  Terms of this form
will turn out to be ``$F$-terms'' when $L\le 2$.

The powers  of momenta, $2\gamma_L$  and $2\beta_L$ in the  low energy
amplitude depend  crucially on the  analysis of the  integrations over
fermionic zero modes.

\subsec{Zero-mode integrals and momentum prefactors}\subseclab\seczeromode

We will now discuss the integration over the fermionic zero modes that
need to be saturated in order to obtain a non-zero contribution to the
amplitude.   As   explained  in~\refs{\BerkovitsVI,\BerkovitsVC},  the
regulator  of~\Reg\  regularizes   divergences  in  the  pure  spinor
functional  integral coming from  $\lambda\,, \bar\lambda  \to \infty$
and in the process  provides essential fermionic zero modes.  However,
to   regularise    potential   divergences   from    the   $\lambda\,,
\bar\lambda\to  0$  endpoint, one  needs  to  introduce  a   small
$\lambda$ regulator which is more complicated
 because it involves non-zero modes of the world-sheet fields.
Fortunately,  it  was  shown in~\refs{\BerkovitsVI,\BerkovitsVC}  that
this  more   complicated  regulator  is   unnecessary  for  evaluating
contributions to  ``$F$-terms'' in the effective action.  Here, we are
defining ``$F$-terms'' as  any term in the effective  action where the
external vertex  operators contribute fewer  than sixteen $\t$  zero modes.
In  other words,  at  least one  $\t$  zero mode  must  come from  the
regulator of~\Reg\ when evaluating an ``$F$-term''.

To absorb the $16L$ zero modes  of $d$ in the most efficient manner in
an $L$-loop  amplitude, the $(3L-3)$ $b$-ghosts  should contribute the
terms  $(\bar\l  \Pi  d)^{L-2}   (\bar\l  r  d^2)^{2L-1}$.  Note  that
increasing the  relative number of  $(\bar\l r d^2)$ terms  will allow
some  $d$'s to  contribute nonzero  modes.  But since  each such  term
includes an extra $r$, it will  increase the number of $\t$ zero modes
which come  from the external  vertex operators. So changing  a $d_\a$
and  $\t^\beta$ zero  mode to  a  $d_\a$ and  $\t^\beta$ nonzero  mode
forces the external vertex operators to contribute two extra $\t$ zero
modes, which  is equivalent  to adding a  factor of momentum.  So each
contraction of $d_\a$ with $\t^\beta$ in the computation adds a factor
of momentum to the term in the effective action.

We now wish to isolate the terms of lowest dimension -- in other words, the terms with the least number of $\theta$'s taken out of the vertex operators. 
Using the normal
coordinate expansion of  \edA\  it is easy to see that the vertex operator  of lowest dimension is the superfield $W^\a$.     Using the above contribution from the $b$-ghosts, the term with the 
lowest power of momentum   in the $L$-loop amplitude (for $L>1$) is proportional to the correlation function
\eqn\eaz{
\int d^{3(L-1)}\tau\,\Big\langle 
\left(r\cdot \theta\right)^{12-2L} 
\left(S \l d\right)^{11L} \,
\left(\bar \l\Pi d\right)^{L-2}
\left(\bar \l \, r\, d^2\right)^{2L-1} \, (\l \bar\l)^{-5L+4}\, (W\, d)^{4} 
\Big\rangle\,.
}
In this expression the first two factors come from expanding the regulator $\calN$, the subsequent four factors come from the $3L-3$ powers of the $b$-ghost, and the last factor comes from the vertex operators.
We see that there are $12-2L$ powers of $\theta$ and therefore the term of lowest dimension at $L$ loops is proportional to
\eqn\ethet{
\int d^{16}\t \, \t^{12-2L}\, W^4 \sim
\p^Lt_8 F^4\,.
}
This expression is symbolic since we have not specified the way in which the gauge indices are contracted or the details of how the derivatives act on the four fields, but these are determined by the explicit calculations.
Since only terms with an even number of momenta
can be non-vanishing, one finds that 
$\p^2t_8 F^4$ is the term of lowest dimension at $L=2$,
$\p^4t_8  F^4$ is the term of lowest dimension at $L=3$ and $L=4$,
and $\p^6t_8  F^4$ is the term of lowest dimension for $L\geq 5$.

This expression suggests  that the  $D$-term of  lowest  dimension is
$\int d^{16}\t  \, W^4\sim \partial^6 F^4$. However,  this is too naive since it assumes that
the remaining integrations over  the non-zero modes in~\eaz\ do  not contribute  inverse derivative  factors  such as
$(k_r\cdot k_s)^{-1}$.   Such factors do arise and play an important r\^ole in determining which terms are genuine $F$-terms and which are fake $F$-terms. The systematics of these inverse derivatives is the subject of the next subsection.

We will also consider the  extension of this argument to four-graviton
amplitudes in  the  type II  closed  string  theory  where it  was  argued
in~\refs{\BerkovitsVC} that terms  proportional to $\partial^{2k} R^4$
with $k\le 5$ are $F$-terms that do not receive corrections beyond $k$
loops.  This  also assumed the  absence of inverse powers  of momentum
arising from non-zero  modes, which we will justify at the end of the
next subsection  for terms that do not require the complicated small $\lambda$ 
regulator.

\subsec{Inverse derivative factors}\subseclab\secmomenta

When computing  a massless four-point $L$-loop  amplitude, one expects
to  get inverse derivative  factors of  $(k_1 \cdot  k_2)^{-1}$ coming
from massless poles  when $(k_1 + k_2)^2 =2k_1\cdot k_2  = 0$. But the
on-shell massless  three-point amplitude vanishes beyond  tree level
(i.e., for  $L\geq 1$) in superstring  theory, so the massless four-point
loop  amplitude  cannot  have  a  physical  pole  when  $k_1\cdot  k_2
=0$. Nevertheless,  there is the  possibility that inverse  factors of
$(k_1\cdot k_2)^{-1}$ could cancel factors of $(k_1\cdot k_2)$ in the
prefactor  of  the  amplitude  from  the zero  mode  saturation. As
will now be discussed, these inverse factors can come from 
performing the contractions over  the non-zero modes of the world-sheet
fields and integrating over the moduli of the amplitude.  This could in principle reduce the
$D$-term of lowest dimension from $\int  d^{16}\t \ W^4 = \p^6 F^4$ to a
term with fewer derivatives.

In the superstring computation, these inverse derivative factors arise
from  the boundary  of moduli  space where either two vertex operators
collide or where the string  world-sheet splits
into  two world-sheets connected by  a long open  string strip (or
closed string  tube). For example,  a factor of  $(k_1\cdot k_2)^{-1}$
could arise from the region of the 
integral $\int dz_2 V_2(z_2) V_1(z_1)$ when $z_2$
approaches $z_1$. This inverse derivative factor occurs if
$V_2(z_2) V_1(z_1)$ has a term in its OPE which goes like
$(z_2-z_1)^{-1 + k_1\cdot k_2}$ so that
$\int dz_2 V_2(z_2) V_1(z_1)
\sim\int dz\ z^{-1  + k_1\cdot  k_2}\sim  (k_1\cdot k_2)^{-1}$.
Similarly, 
a factor of  $(k_1\cdot k_2)^{-1}$
could  arise  from  the   limit  in  which  the  $L$-loop  world-sheet
degenerates into $L_1$-loop and $L_2$-loop world-sheets connected by a
long open string strip (or closed string tube) where $L=L_1 + L_2$ and
$k_1 + k_2$  is the momentum going through the  strip (or tube).  Such
an inverse  derivative factor  could come from  the $y\to 0$
region of the integral  $\int dy
~y^{-1  + k_1\cdot  k_2}\sim  (k_1\cdot k_2)^{-1}$  where  $- \log ( y)$ is  the
length of  the open  string strip,  or from the $y\to 0$
region of the integral $\int  d^2 y
|y|^{-2+  k_1\cdot  k_2}\sim (k_1\cdot  k_2)^{-1}$  where  $- \log( y)$ is  the
complex modulus of the closed string tube.

It will now be shown that when $L<5$, the only possible source of
inverse derivative factors comes from the collision of vertex operators.
Furthermore, these inverse derivative factors only affect the low-energy
dependence of
the $s^{\gamma_L}\,t_8\, \tr(F^4)$ term, and do not affect the $s^{\beta_L}\, t_8(\tr F^2)^2$ term. For $L\geq 5$,
one can also get inverse derivative factors from the degeneration of the
surface, and these inverse factors affect both the $s^{\gamma_L}\,t_8\,\tr(F^4)$ and
$s^{\beta_L}\, t_8(\tr F^2)^2$ terms. It will also be shown that neither of these two sources
of inverse derivative factors affect the low-energy dependence of the
$s^k\, R^4$ terms in closed superstring scattering (at least for terms which do
not involve the small $\lambda$ regulator).

We shall first discuss possible inverse derivative factors coming
from the collision of vertex operators.
As explained in subsection~\seczeromode, the term of fewest derivatives
in the effective action comes if each of the four vertex operators
contributes $W^\alpha d_\a$ and these four $d_\a$'s only contribute zero modes.
When two such vertex operators collide, e.g. $V_1(z_1)$ and $V_2(z_2)$,
the resulting OPE is simply $W_1^\a d_\a W_2^\b d_\b (z_2-z_1)^{k_1\cdot k_2}$.
But to get an inverse derivative factor of $(k_1\cdot k_2)^{-1}$,
the OPE must have a term proportional to $(z_2-z_1)^{-1+k_1\cdot k_2}$. To
get this additional factor of $(z_2-z_1)^{-1}$,
the $d_\a$ variable in one of the vertex operators
must contribute a nonzero mode which contracts with a $\t^\a$
variable in the other vertex operator. 

Note that switching the order
of the two vertex operators will reverse the sign from $(z_2-z_1)^{-1}$
to $(z_1-z_2)^{-1}$, so the resulting OPE is antisymmetric under
exchange of the group theory factors $T_1$ and $T_2$. This immediately implies
that this type of inverse derivative factor, which comes from colliding
vertex operators, is not present for
the $t_8(\tr F^2)^2$ term. If $V_1$ and $V_2$ are on the same boundary for
the double-trace term, $\tr(T_1 T_2) = \tr(T_2 T_1)$ implies that
the antisymmetric part of the OPE does not contribute. This
is related to the fact that the gluon vertex operator
coming from the pole in the OPE of the two external vertex operators 
would be a $U(1)$ gluon which decouples from non-abelian states.

For the $t_8\,\tr(F^4)$ term, these colliding vertex operators could potentially
reduce the number of derivatives. However, for this to happen,
the missing $d_\a$ zero mode from
the colliding
vertex operator needs to be replaced by an extra $d_\a$ zero mode
coming from the $b$ ghosts. The lowest value of $L$ for which this
is possible is $L=3$, as can be seen from \eaz\ -- by changing
the $(\bar\l \Pi d)$
contribution to a $(\bar\l r d^2)$  contribution, one gets an extra
$d_\a$ zero mode. However, because one also gets an extra $r_\a$
zero mode and because one of the $\t$ zero modes in the vertex operators
was contracted with
the $d_\a$ nonzero mode, the number of $\t$'s 
at $L=3$ coming from the vertex operators is increased from 10 to 12. After
including the inverse derivative factor of $(k_1\cdot k_2)^{-1}$, this means that 
the term with fewest derivatives at $L=3$ is
\eqn\eRed{
(k_1\cdot k_2)^{-1} ({\p\over{\p\t}})^{12} W^4 \sim \p^2 \,t_8\tr(F^4)\,.}

If in addition to $V_1$ and $V_2$ colliding, one also had $V_3$ and
$V_4$ colliding, 
one could potentially get an additional inverse
derivative factor of $(k_1\cdot k_2)^{-2}$. In this case, one would
need to get two extra $d_\a$ zero modes from the $b$ ghosts in order
to replace the two $d_\a$ zero modes in the vertex operators which
were contracted with $\t$'s. It is easy to see from \eaz\ that this
is possible at $L=4$ by changing two $(\bar\l\Pi d)$ contributions to
$(\bar\l r d^2)$ contributions. This gives two extra $r_\a$ zero modes and,
because two $\t$ zero modes in the vertex operators are contracted, the
number of $\t$'s at $L=4$ coming from the vertex operators is increased from
12 to 16. After
including the inverse derivative factor of $(k_1\cdot k_2)^{-2}$, this means that 
the term with fewest derivatives at $L=4$ is
\eqn\eRedII{
(k_1\cdot k_2)^{-2} ({\p\over{\p\t}})^{16} W^4 \sim \p^2\,t_8 \tr(F^4)\,,}
resulting in the same operator as in \eRed.

So we have seen that colliding vertex operators reduce the momentum dependence
of the $t_8\,\tr(F^4)$ term from $\p^3 t_8\,\tr(F^4)$ to $\p^2 t_8\,\tr(F^4)$ at
$L=3$, and from $\p^4 t_8\,\tr(F^4)$ to $\p^2 t_8\,\tr(F^4)$ at $L=4$. However,
the momentum dependence of the double-trace term is unaffected and
remains $\p^{2\lceil 3/2\rceil} t_8\,(\tr F^2)^2= \p^4 t_8\,(\tr F^2)^2$ at $L=3$ and $\p^4 t_8\,(\tr F^2)^2$ at $L=4$.

We will now analyze the second possible source of inverse derivative
factors coming from the degeneration of the  $L=L_1+L_2$ world-sheet into two world-sheets with $L_1$ and $L_2$ loops.   In practice we will find it convenient to describe the closed
string version of this plumbing decomposition, which is related
to the open string version by the usual doubling trick.  So we will consider the degeneration of a genus $g=g_1+g_2$
closed-string world-sheet into two world-sheets of genus $g_1$ and $g_2$.
To analyze this degeneration, it is convenient to use the
standard ``plumbing'' decomposition of the surface into a
genus $g_1$ surface with a small hole at $p_1$, a genus $g_2$
surface with a small hole at $p_2$, and a cylinder of length
$-\log(y)$ connecting the two holes.
The Beltrami differential
for the cylinder length is $\int dy y^{-1}$, and the corresponding
$b$ ghost is integrated around the circumference of the cylinder.
The remaining $3g-4$ $b$ ghosts are split into $3g_1 -2$ $b$ ghosts
on the $g_1$ surface (with one of them at $p_1$) and $3g_2-2$
$b$ ghosts on the $g_2$ surface (with one of them at $p_2$).

To compute the correlation function in this degeneration limit, it
is convenient to map the cylinder of length $-\log(y)$
into an annulus which has one small
boundary of radius $\sqrt{y}$ and one large boundary of radius
$1/\sqrt{y}$. The annulus coordinate will be called $w$, which is related
to the cylinder coordinate $\rho$ by $w=e^\rho$.
To get an inverse derivative factor, the correlation function on
the annulus must contribute a factor of
$y^{k_1\cdot k_2}$, so that after integrating over the moduli
using the Beltrami differential for $y$, one
gets a factor of $\int dy y^{-1} y^{k_1\cdot k_2}\sim (k_1\cdot k_2)^{-1}.$

The correlation function on the annulus is given by 
\eqn\eBB{\langle V_1(w=\sqrt{y}) ~\int dw w b(w) ~V_2(w=\sqrt{1/y})\rangle}
where $V_1$ comes from the operators on the genus $g_1$ surface,
$V_2$ comes from the operators on the genus $g_2$ surface, and
the $b$ ghost is integrated around the countour $|w|=1$. 
If gluons 1 and 2 are on the $g_1$ surface and gluons 3 and 4 are on
the $g_2$ surface, $V_1$ will be proportional to $e^{i(k_1+k_2)x}$ and
$V_2$ will be proportional to $e^{-i(k_1+k_2)x}$. So if the $b$ ghost
does not contribute any $y$ dependence, the correlation
function gives the desired factor of $y^{k_1\cdot k_2}$.

However, when $y\to 0$, the $b$ ghost cannot contribute any $d_\a$ or $\Pi$
zero modes. This is easy to see since when $y\to 0$, the $g$ holomorphic
one-forms split into
$g_1$ holomorphic one-forms which are non-vanishing only on the genus
$g_1$ surface and $g_2$ holomorphic one-forms which are non-vanishing
only on the genus $g_2$ surface.  The $b$ ghost on the annulus contributes
either the term $\bar\lambda\Pi d$ or $\bar\l r d^2$. In the first
case, the nonzero mode of the $\Pi$ must contract with $e^{i(k_1+k_2)x}$
to give a factor of $k$, and the nonzero mode of $d_\a$ must contract with
a $\t^\a$ in $V_1$ or $V_2$. Furthermore, one needs an extra $d_\a$ zero
mode to come from one of the other $b$ ghosts (which is possible when
$L\geq 3$ as before). Putting all of these factors together, one gets
a total factor of $k^2$ which cancels the inverse
derivative factor of $(k_1\cdot k_2)^{-1}$. 
In the second case where the $b$ ghost on the annulus contributes $\bar\l
r d^2$, the two nonzero modes of $d_\a$ must contract with two $\t^\a$'s
in $V_1$ or $V_2$, and one needs two extra $d_\a$ zero modes to come
from the other $b$ ghosts (which is possible when $L\geq 4$). Again putting
these factors together, one gets a total factor of $k^2$ which cancels
the inverse derivative factor of $(k_1\cdot k_2)^{-1}$.

So degeneration of the surface cannot give inverse
derivative factors when $L<5$. However, when $L\geq 5$, there are 
contributions to the low-energy effective action which require
using the more complicated  small $\lambda$ regulator. Although we will not go into
details here, we will sketch how degeneration of the surface can
give rise to inverse derivative factors when $L\geq 5$.

When $L=5$, there are 12 $b$ ghosts which can contribute the term
$(\bar\l r d^2)^{12} \sim r^{12} d^{24}$. This term diverges when $\l\to 0$ as
$\l^{-12}$ which means one needs to use the complicated version
of the regulator. Note that one of the 12 $r$'s in this term must
contribute a nonzero mode (since there are only 11 independent $r_\a$
zero modes). As shown in \refs{\BerkovitsVI,\AisakaYP}, the 
 small $\lambda$ regulator involves a term
$d_\alpha s^\alpha$ with nonzero modes, and after contracting the $s$ nonzero
mode in the regulator
with the $r_\a$ nonzero mode, one is left with a term proportional to
$r^{11} d^{25}$. After including the 4 $d_\a$'s from the vertex operators
and the $5\times 11=55$ $d_\a$ zero modes from the regulator, 
one has a total of 84 $d_\a$'s. At $L=5$, one needs 80 $d_\a$ zero modes,
so 4 of the 84 $d_\a$'s can contribute nonzero modes. 

If the $L=5$ surface degenerates in two places, one gets two inverse
derivative factors which produce $(k_1\cdot k_2)^{-2}$. 
The two $b$ ghosts on the two
annuli will involve these four $d_\a$ nonzero modes which will contract
with four of the $\t$'s on $V_1$ or $V_2$. Since no $\t$'s come
from the regulator, one needs to get 20 $\t$'s
from the vertex operators. So at $L=5$, the number of derivatives in
the double-trace term is
$k^{-4} ({\p\over{\p\t}})^{20} \tr(W^2)^2 \sim \p^4\,t_8( \tr F^2)^2$.
Note that this is one derivative lower than
the $\p^5 t_8\,(\tr F^2)^2$ term one would get in the absence of inverse derivative
factors. 

For the single-trace term, one can use two of these four $d_\a$ nonzero modes
to give an inverse derivative factor of $k^{-4}$ from colliding vertex 
operators $V_1$ with $V_2$ and $V_3$ with $V_4$. The remaining two $d_\a$
nonzero modes can be used in the $b$ ghost on an
annulus which degenerates the $L=5$ surface in one place, which gives
an additional factor of $k^{-2}$. Since each
$d_\a$ nonzero mode is contracted with a $\t$ from the vertex operators,
one again needs 20 $\t$'s from the vertex operators,
so at $L=5$ the single-trace term is proportional to
$k^{-6} ({\p\over{\p\t}})^{20} \tr(W^4) \sim \p^2 \,t_8\tr (F^4)$. 

So after including the inverse derivative factors coming from these two
sources, one finds that the single-trace term is proportional to
$\p^2 t_8\,\tr(F^4)$ for $L>1$ and the double-trace term is proportional
to $\p^2 t_8\,(\tr F^2)^2$ for $1<L\leq 2$ and $\p^4 t_8\,(\tr F^2)^2$ for $L>2$.
This was verified up to $L=5$.

Finally, it is easy to show that  for terms which do not involve
the small $\lambda$ regulator, neither of these sources 
can contribute inverse derivative factors for closed string scattering. This
is because the number of $d_\alpha$ and $\t^\a$ nonzero modes
is doubled (since one has both left and right-moving contributions),
but the number of inverse derivative
factors from the massless propagator $(k_1\cdot k_2)^{-1}$ is the same as in the open string scattering.
So the number of $k$'s in the numerator is always equal or greater
than the number of $k$'s in the denominator.
 For terms involving the small $\lambda$ regulator, the analysis of
inverse derivative factors is more complicated and will not be
attempted in this paper.

The  previous  paragraph  clarifies  a statement  in\refs{\BerkovitsVC}
concerning  graviton scattering  in type  II theories.  In  that case,
integration  over fermionic  and bosonic  zero modes  in four-graviton
scattering at  genus $g$ led  to a prefactor  in the low-energy amplitude  of the
form $s^{g}  R^4\, I(s,t,u)$ for  $g\le 6$, multiplying  a complicated
dynamical function of the external momenta.  The conclusion that these
terms are $F$-terms relied on the  absence of inverse powers of $s$ in
$I(s,t,u)$ arising from potential closed-string poles in the dynamical
factor  multiplying the  zero-mode  prefactor.  The  arguments of  the
previous paragraph verify that no  such inverse powers are present and
these terms are indeed $F$-terms,  at least up to $L=5$ where
 the complicated small $\lambda$ regulator is unnecessary.  This lends support to the arguments
in~\refs{\GreenYU} that ultraviolet  divergences are absent in $\cN=8$
supergravity up to at least eight loops.

\newsec{Contributions of handles}\seclab\sechandles

Open superstring theory is well-known to generate a closed-string, or gravitational, sector in string perturbation theory, starting at one loop ($L=1$).   
In order to decouple gravity in the low energy limit  one needs to take the large-$N$ limit, where each additional handle is suppressed by a power of $1/N^2$.  Nevertheless,
it is interesting to consider how the inclusion of handles in the 
world-sheet computation at finite $N$ (and the resulting coupling to closed string modes)
affects the $t_8\tr(F^4)$ and $t_8(\tr F^2)^2$ terms in the low-energy 
superstring effective action.

We will now suppose the world-sheet has $H$ handles and $B$ boundaries.
The open string amplitude with $M$ vertex operators on a world-sheet of given topology 
carries a factor
 $g_s^{-\chi}$, where  $\chi=2-2H-B$ is the Euler  number (where we again set the  number of crosscaps  $C=0$ as appropriate for oriented
strings). We are particularly  interested in the case $M=4$.  
Note that  each handle  is associated with  a power $g_s^2$,  which is
equivalent  to  the insertion  of  two  boundaries,  but whereas  each
free boundary generates a factor $N$, a handle does not depend on $N$.
Each handle  is associated with  three complex moduli and  bring three
$b$-ghosts  and their  complex conjugates, so  a world-sheet  with $H$
handles and  $B$ boundaries has  $3(2H+B-2)$ $b$-ghost insertions.
The four-gluon amplitude on  a world-sheet with $L=B+2H-1$ boundaries and
$H$ handles  is a  simple generalization of~\eAmpone\ and~\eAmp, with
each handle  counting as two  boundaries and an appropriate  change in
the power of $N$ as described above.  

\fig{
Three-loop contribution to the single-trace term in four-gluon scattering and its connection to the $\tr F^4$ term in Yang--Mills theory in the low energy limit.}
{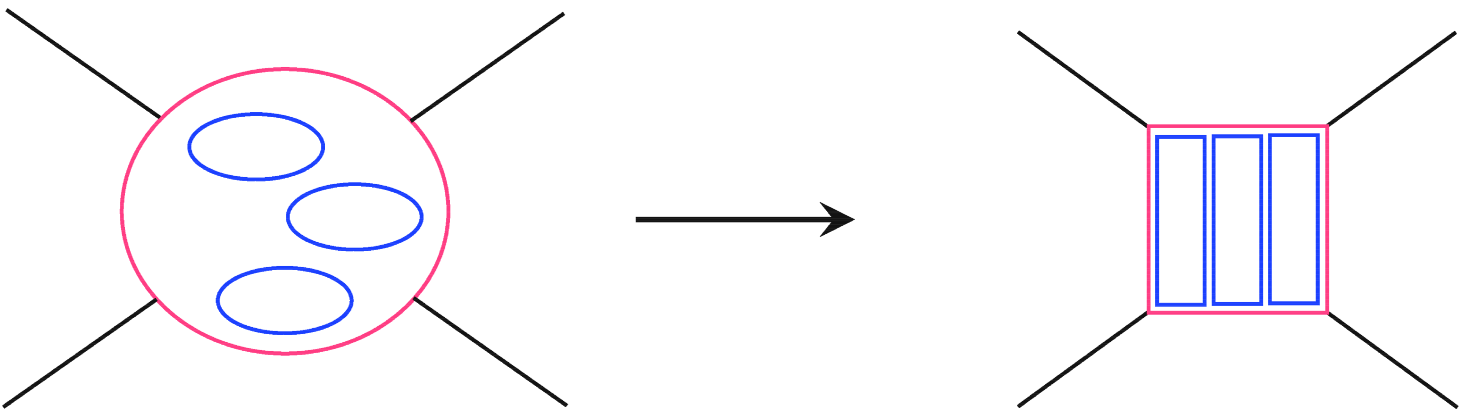}{10.cm}
\figlabel\planar
\fig{
A world-sheet with one handle and two boundaries that contributes to the single-trace term at the same order as the four-boundary amplitude of figure~\planar.  The figure illustrates examples of field theory diagrams that arise from the short-handle and long-handle boundaries of moduli space  in the low energy limit.}
{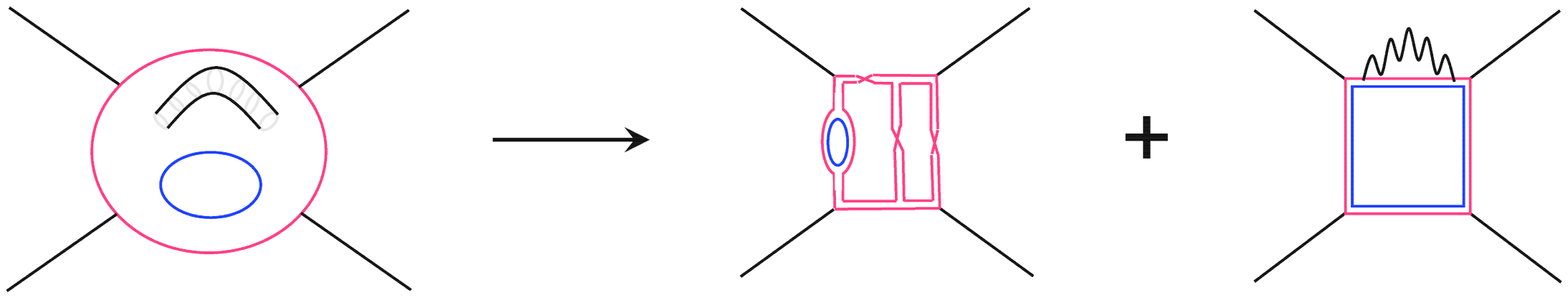}{12.cm}
\figlabel\nonplanar

 Figure~\planar\ illustrates a contribution to the single-trace amplitude with three loops and no handles, $H=0, B=4$, and an example of a Feynman diagram that arises in the low energy planar Yang--Mills limit.  Figure~\nonplanar\ illustrates a contribution of the same order in string coupling that has one handle, $H=1, B=2$.   Two distinct boundaries of moduli space contribute in the  low energy field theory limit.  One of
these arises from  the short handle limit, which  contributes a sum of
three-loop  non-planar Yang--Mills  diagrams, which  give contributions
analogous to  those of  figure~\planar, but with  two fewer  powers of
$N$.  The other arises from the  limit in which the handle is long and
picks out  the ground state graviton  exchange, resulting in  a sum of
Yang--Mills loop diagrams with a graviton propagator attached.

The  first  example where a handle contributes
  is  the  addition  of a  handle  ($H=1$)  to  the
tree-level disk  world-sheet -- i.e.,  to the world-sheet  with $B=1$.
This is associated with a factor  of $g_s\, s\, t_8\tr F^4$ in the low
energy limit, whereas the  two-loop ($B=3$, $H=0$) $t_8\tr\, F^4$ term
has a  factor of  $g_s\, N^2\, s\, t_8 \tr F^4$.  So to order  $g_s$, the
analytic contribution  to the  low-energy expansion of  the four-point
open string amplitude is given by
\eqn\eTwoALLN{
 \cA^{ana.}_{L=2}\sim g_s \ \, \left((c_{3,0} N^2 + c_{1,1}) \, s\, t_8\, \tr(F^4) + d_{3,0} N
s\, t_8(\tr F^2)^2\right) \ , }
where $c_{B,H}$ and $d_{B,H}$ are coefficients which in principle
could be computed.
There  is no
canonical way  of separating the open-string  and closed contributions
to the $c_{1,1}$ coefficient which is subleading in $1/N^2$.  
We will  return  to this point  in the  next
section  when we  will  discuss  the connection  with  the pure  field
theoretical super-Yang-Mills results of~\refs{\BernCT,\DixonTalk,\BernNH}.
 
 With  $L=3$  the $B=4$, $H=0$ term is accompanied by a contribution with
$B=2$, $H=1$, which is a $1/N^2$ correction.  In this case the leading
contribution to the 
analytic part of  the low-energy expansion of the three-loop four-point open string amplitude is given by
 \eqn\eThreeALLN{
\cA_3^{ana.}\sim g_s^2\, \left(  ( c_{4,0} N^3 + c_{2,1}\, N) \,
s\, t_8\, \tr(F^4) +
(d_{4,0} N^2 + d_{2,1})s^2\, t_8(\tr F^2)^2\right) 
\ , }
where the double-trace term is proportional to $s^2$ because of the
absence of inverse derivative factors as
described earlier in section~\secmomenta.

With  $L=4$    the  $B=5$, $H=0$  term  is accompanied  by  the
contributions of a world-sheet with $B=3$, $H=1$ and $B=1$, $L=2$ for both the single trace and the double trace terms.   In this case the leading
contribution to the 
analytic part of  the low-energy expansion of the three-loop four-point open string amplitude is given by
\eqn\eFourALLN{
\cA_4^{ana.}\sim g_s^3\, \left((c_{5,0} N^4 + c_{3,1}\, N^2+c_{1,2}) \,
s\, t_8\, \tr(F^4)+ 
(d_{5,0} N^3 + d_{3,1} N)s^2\, t_8(\tr F^2)^2\right) 
\,.
 }
%

\newsec{Connections with maximal SYM theory in various dimensions}
 
 We will now discuss the connections between the low energy limit of open-string results with multi-loop amplitudes in  maximally supersymmetric  Yang--Mills  field theory in $D$ dimensions.
We  are   considering  the  open  string  theory   on  $N$  coincident
$Dp$-branes where $p=D-1$.   If  the  corresponding  low energy  $U(N)$  Yang--Mills
theory coupling constant (which  has non-zero dimension when $D>4$) is
fixed so that 
\eqn\eSYM{
g^2_{\rm YM } = g_s\, l_s^{D-4}= {\rm constant}
}
as $l_s\to 0$, the gravitational coupling vanishes,
\eqn\kappadef{
\kappa^2 = g_s^2\, l_s^8 = g_{\rm YM}^4\, l_s^{16-2D} \to 0\, ,
}  
provided $D<8$ (or $p<7$), which is the condition that the gravitational back reaction of the $Dp$-brane can be ignored.   

As  mentioned earlier,  although we  are decoupling  the closed-string
sector,  world-sheet  handles  are  nevertheless expected  to  make  a
contribution to the theory in the low energy limit. 
In section~\sechandles\ we  discussed the effect  of handles
on the  low-energy expansion  of the open  string amplitudes and we saw
that there  is  no  way of  separating  `open  string'
contributions from `closed  string' contributions to the sub-leading
$1/N^2$ corrections.
One  way  to  see  this  is  to  consider,  for  instance,  the  following
contribution  to   the  two-loop  effective   action  from  the
four-point amplitude
\eqn\eL{
S_{L=2} = \int d^Dx \, \sqrt{-g}\, g_s \, l_s^{10-D} \, (c_{3,0} N^2 + c_{1,1})
\, \partial^2 t_8\tr(F^4).
} 
Using the relations~\eSYM\ and~\kappadef\ one can write the $1/N^2$ correction
to  the effective  action  either as  a super-Yang-Mills  contribution
$c_{1,1}\, g_{\rm YM}^2\, l_s^{2(7-D)}\partial^2 t_8\tr(F^4)$ or as a mixed Yang-Mills
and gravity contribution $c_{1,1}\,\kappa\,l_s^{6-D}\, \partial^2 t_8\tr(F^4)$.
Because there is no way of (and no meaning in) separating the gravitational contribution
from the super-Yang-Mills in string  theory, we will only focus on the
large $N$ contribution  by restricting our attention  to the terms  of order $N^L$
and $N^{L-1}$, which get no contribution from world-sheet handles.

\subsec{Onset of ultraviolet divergences in various  dimensions}

In the  limit $l_s\to 0$ the  string theory results have clear implications for the structure of the ultraviolet divergences of  the four-gluon SYM field theory
calculations.  In  particular, we  see why the  single trace  term has
worse ultraviolet divergences than the double-trace term.  A clear way
of  characterising this  is by  determining the  ``critical dimension'', $D_c$, which is the  minimal  dimension in
which  a  $L$-loop term  diverges  in  the  ultraviolet --  i.e.,  the
dimension in which the  ultraviolet divergence is logarithmic.

To begin, we note that the superficial degree of divergence of a $L$-loop 
 Feynman diagram contribution to the four-gluon amplitude in $D$ dimensions is $\Lambda^{(D-4)L}$,
 where  $\Lambda$  is a  momentum  cut-off.   Since  there is  also  a
 prefactor of $t_8 F^4$ at all orders, the divergence is reduced to $\Lambda^{(D-4)L-4}$.  However, in
the case of the single trace term we found that there is a factor of $g_s^{L-1}\, \p^2\, t_8\tr F^4$ for  $1<L\le 5$, so that the degree of divergence is  $\Lambda^{(D-4)L-6}$.  We therefore reproduce the result that the single-trace term is ultraviolet finite in  dimensions satisfying $D <D_c = 4+6/L$,
at  least up to  $L=5$ and  quite probably  for all  $L$.
In these
dimensions the amplitude has a negative mass dimension indicating the presence of infrared divergences given by inverse
powers of the external momenta.
In the case of the $L$-loop contribution to the double-trace term the prefactor has the form 
$g_s^{L-1}\,  \p^{2\lceil L/2  \rceil}\, t_8(\tr  F^2)^2$ so  that the
degree of divergence is, for $1<L\le 4$,
 $\Lambda^{(D-4)L-2 \lceil L/2 \rceil - 4}$. For this range of $L$ the amplitude is ultraviolet finite in dimensions satisfying 
 $D< D_c = 4 +(4 + 2\lceil L/2 \rceil)  /L$.
Although we have no firm statements at higher loops, we expect that
since $\partial^4\, t_8(\tr F^2)^2$ is a $D$-term it will receive corrections
from all $L\ge 5$. It would  then follows that for $L\geq 3$ there are
no ultraviolet divergences in $D< D_c=4+8/L$.  In these dimensions the double trace contribution
to the amplitude has a negative mass dimension, again indicating the presence of infrared divergences represented by inverse
powers of the external momenta.
 The results  are summarised  by the table  below and match  the field
 theory    results   of~\refs{\BernCT,\DixonTalk,\BernNH}    for   the
 evaluation of the four gluon amplitude up to $L=4$.
 
\bigskip

\hbox to \hsize{\hfill\vbox{
\offinterlineskip
\tabskip=0pt
\halign{ 
\vrule height2.75ex depth1.25ex width 0.6pt $#$\tabskip=1em &
&\vrule\     \hfil$#$\hfil\    &\vrule\     \hfil$#$\hfil\    &\vrule\
\hfil$#$\hfil\  &\vrule\
\hfil$#$\hfil\ 
&\vrule\ \hfil$#$\hfil \vrule width 0.6pt \tabskip=0pt\cr
\noalign{\hrule height 0.6pt}
&L=1&L=2&L=3&L=4&L=5\cr
\noalign{\hrule }
\ \p^{2\gamma_L}\,t_8\tr(F^4)&D_c=8& D_c=7&D_c=6&D_c=11/2&D_c=26/5\ \cr
&\gamma_1=0&\gamma_2=1&\gamma_3=1&\gamma_4=1&\omit    \vrule   \
\hfil $\gamma_5=1$ \hfil\vrule\cr
\noalign{\hrule height 0.6pt}
 \ \p^{2\beta_L}\,t_8(\tr F^2)^2&D_c=8&D_c=7&D_c=20/3&D_c=6&D_c=28/5\cr
&\beta_1=0&\beta_2=1&\beta_3=2&\beta_4=2&\omit    \vrule   \
\hfil $\beta_5=2$ \hfil\vrule\cr
\noalign{\hrule height 0.6pt}
}}\hfill}

\bigskip

 Although there is no canonical way of
separating  the gravitational contributions  from the  pure Yang-Mills
corrections in string theory, the $1/N^2$ corrections described in section~\sechandles\
qualitatively  reproduce the  result quoted  in~\refs{\DixonTalk} with
the exception of  the absence of a contribution  independent of $N$ to
the $\p^2 \,t_8\tr F^4$ counterterm  at four loops in $D=11/2$.

\newsec{Summary and comments on higher-point amplitudes}

In this paper, we have analyzed open superstring four-point amplitudes using
the pure spinor formalism and determined non-renormalization
properties of certain terms in the low-energy
effective action. Terms in the effective action proportional to
$t_8 \tr (F^4)$ and $t_8 (\tr F^2)^2$ were shown not to receive corrections
above one-loop, as expected from their connection to the anomaly-cancelling
term $B\wedge F^4$. 
Furthermore, the $\partial^2 t_8 (\tr F^2)^2$ term was shown to not
receive corrections above two loops.
On the other hand,
the $\partial^4 t_8 (\tr F^2)^2$ and
$\partial^2 t_8 \tr (F^4)$ terms are expected to receive corrections
to all loops. These statements were verified up to five loops using
the pure spinor prescription for the four-point amplitudes.

This behaviour can be heuristically explained using
supersymmetry arguments based on $F$-terms and $D$-terms.
The terms $t_8 \tr (F^4)$, $t_8(\tr F^2)^2$ and 
$\partial^2 t_8 (\tr F^2)^2$ are $F$-terms which are expected
to
satisfy non-renormalization conditions.
For $\p^2\, t_8\tr (F^4)$ and $\p^4\,t_8(\tr F^2)^2$ 
the behavior is different since when $D<10$,
$\p^2 \,t_8 \tr(F^4)$ can be written as the $D$-term, $\int d^{16}\theta\ 
\tr (\varphi \varphi)$, where $\varphi$ is a non-linear superfield whose
lowest  component  is a  scalar,  and  $\p^4\,t_8(\tr  F^2)^2$ can  be
 expressed    in   terms    of    $\int   d^{16}\theta    \,
\tr(\varphi\varphi)^2$  and $\int d^4x \int d^{16}\theta
 \,T^2$ where $T_{AB}$ is the symmetric traceless supercurrent~\refs{\DrummondEX}.  So $\p^2 t_8
\tr(F^4)$ and $\p^4\,t_8(\tr F^2)^2$ are not expected
to  satisfy  any   non-renormalization  conditions. 
In the analysis of open superstring amplitudes, the 
$t_8(\tr F^2)^2$ and $t_8\tr (F^4)$ terms behave differently since
inverse derivative
factors from colliding vertex operators
are present in the $t_8\tr(F^4)$ computation but are absent
in the $t_8(\tr F^2)^2$ computation. It would be useful to better understand
the relation between these inverse derivative factors and the nonlinear
construction of the $D$-term $\int d^{16}\theta ~\tr(\varphi\varphi)$.

Our analysis of $t_8\tr(F^4)$ and $t_8(\tr F^2)^2$ terms is consistent with
the field theory computations of 
\refs{\BernCT,\DixonTalk,\BernNH} and explains the apparent puzzle that
$t_8(\tr F^2)^2$ terms are less divergent in the ultraviolet than
$t_8\tr(F^4)$ terms. In addition, our analysis showed that there are
no inverse derivative factors in the analogous Type II computation, 
 at least for terms which do not require the small $\lambda$ regulator.
This lends support to the previous claim of~\refs{\BerkovitsVC}
that for $g<6$, $\partial^{2g} R^4$ terms do
not receive contributions above genus $g$ and to
the   arguments
in~\refs{\GreenYU}  that ultraviolet divergences  are absent  in four-dimensional $\cN=8$
supergravity up to at least eight loops. Therefore, the first ``surprise'' would arise if the four-graviton amplitude was not ultraviolet divergent at nine loops in four dimensions.

It would be very interesting to generalize the methods of this paper
to higher-point amplitudes beyond four points. Since higher-point
amplitudes have massless poles, one needs to first subtract out the
massless poles before using the amplitudes to determine non-renormalization
properties of terms in the low-energy effective action. 
At the moment, it is unclear how to verify that subtracting out
the massless poles does not affect the non-renormalization
properties implied by the zero-mode counting. Nevertheless, one
expects that certain higher-point terms in the effective
action will satisfy non-renormalization conditions and one can
make some preliminary speculations on how the behavior
of $F^4$ terms extend to $F^n$ terms.
In particular,
 the extension of our analysis of the zero mode saturation
to  five-point amplitudes indicates that the 
$\tr F^5$, $\tr F^3\times \tr F^2$ should be one-loop exact, that the 
$\p^2\, \tr F^5$ and $\p^2\,(\tr F^3)( \tr F^2)$ should be two-loop exact,
while
$\p^4\, F^5$  are $D$-terms and  should get contributions to all  loops 
for all group theory structures.  For the six-point  amplitude, the zero mode saturation indicates  
  that the  $\tr F^6$,
$(\tr F^3)^2$ and $(\tr F^2)^3$ should be one-loop exact, while
$\p^2\,  F^6$ should receive  perturbative contributions  to all  orders
for all group theory structures.

\bigskip
\noindent{\bf Acknowledgements:}
We would like to thank Zvi  Bern, Kelly Stelle, Radu Roiban and Edward
Witten for discussions.
The research of NB was partially supported by CNPq grant 300256/94-9
and FAPESP grant 04/11426-0.
The research of PV was supported in part by the ANR grant BLAN06-3-137168.
JGR acknowledges support by research grants MCYT FPA 2007-66665,
2009SGR502.

\footatend\vfill\supereject\immediate\closeout\rfile\writestoppt
\baselineskip=14pt\centerline{{\bf References}}\bigskip{\frenchspacing%
\parindent=20pt\escapechar=` \input refs.tmp\vfill\eject}\nonfrenchspacing
\bye